# Short-Range Polaron Correlations in the Ferromagnetic La$_{1-x}$Ca$_x$MnO$_3$

Pengcheng Dai,[1] J. A. Fernandez-Baca,[1] N. Wakabayashi,[2] E. W. Plummer,[1,3] Y. Tomioka,[4] and Y. Tokura[4,5]

[1]*Solid State Division, Oak Ridge National Laboratory, Oak Ridge, Tennessee 37831-6393*
[2]*Department of Physics, Keio University, 3-14-1 Hiyoshi, Kohoku-ku, Yokohama 223-8522, Japan*
[3]*Department of Physics, University of Tennessee, Knoxville, Tennessee 37966*
[4]*Joint Research Center for Atom Technology (JRCAT), Tsukuba 305-8562, Japan*
[5]*Department of Applied Physics, University of Tokyo, Tokyo 113-8656, Japan*

We use neutron scattering to demonstrate the presence of lattice polarons and their short-range correlations for several samples of La$_{1-x}$Ca$_x$MnO$_3$ in the Ca doping range $0.15 \leq x \leq 0.3$. We establish the doping dependence of the orientation, commensuration, and coherence length of the polaron correlations and show that the populations of correlated and uncorrelated polarons are intimately related to the transport properties of the materials.

The complex interplay between charge, spin, orbital, and lattice degrees of freedom is responsible for the rich phase diagram in the doped perovskite manganese oxides La$_{1-x}A_x$MnO$_3$ (where $A$ is Ca or Sr) [1]. For the electronic hole doping concentration $x \sim 0.3$, these materials exhibit long-range ferromagnetic ordering accompanied by a dramatic increase in electrical conductivity below the Curie temperature $T_C$. The basic microscopic mechanism responsible for this behavior is believed to be the double-exchange (DE) interaction [2], where the hopping of an itinerant $e_g$ electron from the trivalent Mn$^{3+}$ to the tetravalent Mn$^{4+}$ site facilitates both the ferromagnetism and electrical conductivity. It is known, however, that La$_{1-x}$Ca$_x$MnO$_3$ (LCMO) in the doping range $0.15 \leq x \leq 0.22$ has an insulating ferromagnetic ground state [3,4], contrary to the expectation of the DE model. To understand this intriguing behavior, systematic neutron scattering studies of LCMO in the doping range $0.15 \leq x \leq 0.3$ were performed. We observe neutron diffuse scattering arising from lattice distortions associated with polarons [5] and their short-range correlations. As these lattice distortions (polarons) occur only for manganese sites containing $e_g$ electrons (Mn$^{3+}$) [5], short-range polaron-polaron correlations can also be regarded as short-range charge ordering. We establish the doping dependence of the orientation, commensuration, and coherence length of the charge correlations and show that the populations of correlated and uncorrelated polarons are intimately related to the transport properties of these materials irrespective of their insulating or metallic ground state. Our results strongly suggest that the resistivity in these materials is controlled by the competition between short-range charge correlations and long-range ferromagnetic DE interactions.

For this study, we prepared single crystals of LCMO with nominal doping $x \sim 0.15$ (LCMO15, $T_C = 160$ K), 0.2 (LCMO20, $T_C = 178$ K), 0.25 (LCMO25, $T_C = 190$ K), and 0.3 (LCMO30, $T_C = 238$ K) by the floating-zone method [4]. Transport and electron-probe microanalysis on different parts of the crystal indicated that the Ca doping was homogeneous. While LCMO15 and LCMO20 have an insulating ground state, LCMO25 and LCMCO30 exhibit low temperature metallic behavior [4]. All four crystals have mosaic spreads of about 1° and volumes between 0.15 and 0.4 cm$^3$. The experiments were carried out using the HB-1 and HB-1A triple-axis spectrometers, and the wide angle neutron diffractometer (WAND) at the High-Flux Isotope reactor at Oak Ridge National Laboratory. For the triple-axis measurements, we used pyrolytic graphite as the monochromator, analyzer, and filters. The final neutron energy was fixed at $E_f = 13.6$ meV with collimations of, proceeding from the reactor to the detector, 48–40–60–120 min [full width at half maximum (FWHM)]. The WAND diffractometer [6] uses the (3, 1, 1) reflection of a bent silicon monochromator to select an incident beam wavelength of 1.44 Å. The diffracted neutrons are detected by a curved one-dimensional position sensitive $^3$He detector that covers a scattering angle of 125°. LCMO manganites have orthorhombic structures slightly distorted from the cubic lattice. For simplicity, we label the wave vectors $Q = (q_x, q_y, q_z)$ in units of Å$^{-1}$ as $(H, K, L) = (q_x a/2\pi, q_y a/2\pi, q_z a/2\pi)$ in the reciprocal lattice units (rlu) appropriate for the pseudocubic unit cells of LCMO with lattice parameters $a \approx 3.88$, 3.87, and 3.86 Å for LCMO15, LCMO20, and LCMO30, respectively. The crystals were oriented to allow wave vectors of the form $(H, K, 0)$ to be accessible in the scattering plane.

The left panels of Fig. 1 show the temperature dependence of the resistivity for LCMO15, LCMO20, and LCMO30. While LCMO30 exhibits an insulator-to-metal transition below $T_C$ (Fig. 1e), the resistivity of LCMO15 and LCMO20 first decreases below $T_C$ but, upon further cooling, shows a sharp upturn with a low temperature insulating behavior very similar to that of La$_{0.88}$Sr$_{0.12}$MnO$_3$ (LSMO12) [7]. For LSMO12, which is also a ferromagnetic insulator, several microscopic origins have been proposed to explain the mysterious resistivity rise below $T_C$. First, Yamada *et al.* [8] observed a satellite peak at

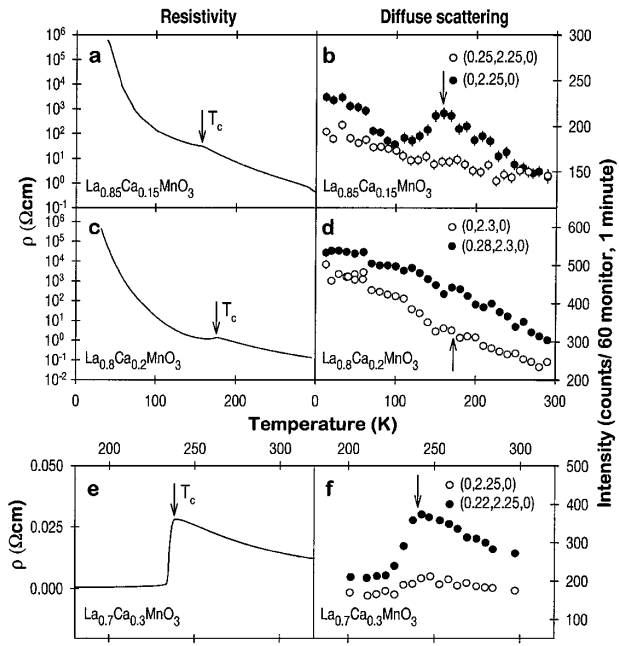

FIG. 1. Temperature dependence of the resistivity for (**a**) LCMO15, (**c**) LCMO20, and (**e**) LCMO30. The intensity of neutron diffuse scattering at the peak positions of charge ordering modulation wave vectors (•) and away from them (∘) for (**b**) LCMO15, (**d**) LCMO20, and (**f**) LCMO30. The error bars include only statistical errors and are given by vertical lines or smaller than the symbol size.

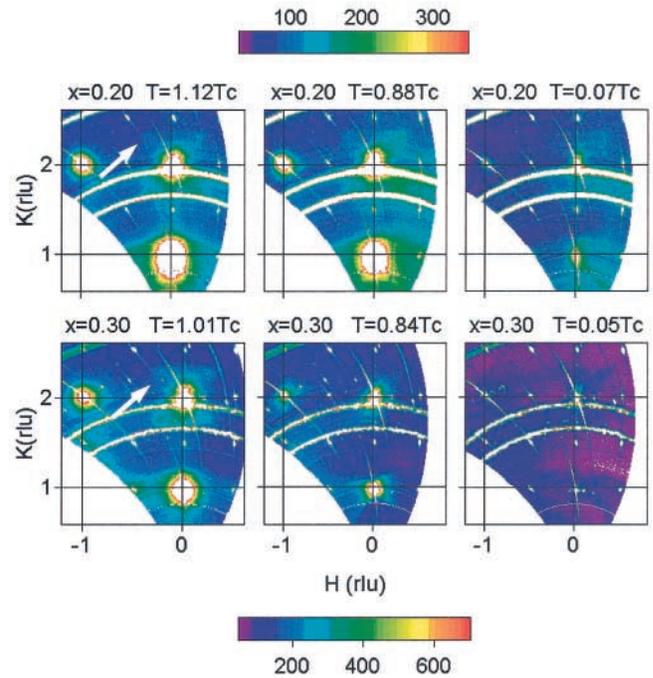

FIG. 2 (color). The observed neutron diffraction patterns of LCMO20 and LCMO30 at various temperatures in the $[H, K, 0]$ reciprocal plane obtained by the WAND diffractometer. The arrows indicate structural diffuse scattering, and the horizontal color bars represent the intensities in counts per 5 min.

$(1/4, 2, 0)$ appearing below the temperature where the resistivity shows the upturn. From the positions of this and other superlattice peaks, the authors inferred a long-range polaron (charge) ordered superlattice structure which has a $2 \times 2 \times 4$ unit cell in terms of the cubic perovskite unit. In this scenario, the resistivity rise below $T_C$ is due to long-range polaron (charge) ordering [8]. Endoh *et al.* [9] confirmed the superlattice reflections of Ref. [8] but found no evidence for long-range polaron ordering. Instead, the authors argue that the resistivity rise in LSMO12 is the consequence of orbital ordering as confirmed by the resonant x-ray scattering. Finally, the origin of the resistivity rise has been attributed to the onset of long-range antiferromagnetic (AF) order in LSMO12 [10], through either canted ferromagnetism [10] or microscopic phase separation [11].

If the resistivity rise in lightly doped LCMO (Figs. 1a and 1c) has the same microscopic origin as in LSMO12, one would expect to observe either superlattice spots at $(1/4, 2, 0)$ and equivalent positions as the signature of polaron (charge) ordering [8] or long-range ordered AF peaks at $(H \pm 1/2, 0, 0)$ [10]. To determine whether the LCMO manganites exhibit such order, we surveyed the reciprocal space using the WAND in the $[H, K, 0]$ reciprocal lattice plane at several temperatures for all four samples. Figure 2 compares the diffraction patterns of LCMO20 and LCMO30 which have ferromagnetic insulating and ferromagnetic metallic ground states, respectively. In addition to the main Bragg reflections from the pseudo cubic structure at $(H, K, 0)$ ($H, K$: integers), we observed satellite reflections associated with lattice distortions from the ideal cubic structure at $(H/2, K/2, 0)$ positions. Although the intensities of these satellite reflections increase monotonically with decreasing temperature, there is no signature of a transition to a canted AF structure [10]. Moreover, since LCMO30 shows metallic behavior below $T_C$ while the lower doping LCMO20 has an insulating ground state (Fig. 1), these superlattice reflections seen in both systems (Fig. 2) must not be associated with polaron or charge ordering. In addition, there is no evidence for the low temperature, charge ordering superlattice peaks at $(1/4, 2, 0)$ or other positions observed in LSMO12 [8].

However, the data in Fig. 2 reveal the clear presence of diffuse scattering around the $(0, 1, 0)$, $(0, 2, 0)$, and $(-1, 2, 0)$ main Bragg peaks which can be either magnetic or structural in nature. The scattering around the $(0, 1, 0)$ and $(-1, 2, 0)$ reflections is isotropic and shows no major change in its characteristics with increasing hole doping. This diffuse intensity decreases in the low temperature ferromagnetic state, consistent with the suppression of the isotropic magnetic diffuse scattering below $T_C$. The diffuse scattering around the $(0, 2, 0)$ reflection, on the other hand, displays distinct butterfly-shaped features (see arrows in Fig. 2) different from the isotropic magnetic diffuse scattering. For LCMO30, this diffuse intensity vanishes in the low temperature metallic state (Fig. 2, the second row). In contrast, the same diffuse scattering increases below $T_C$ for ferromagnetic insulating LCMO20 (Fig. 2, the first row). Thus, the butterfly-shaped diffuse scattering at about $(0,2,0)$ is not magnetic in origin, but the result of lattice distortions. Note that the rings around the origin in Fig. 2 correspond to scattering from the aluminum sample holder, and the narrow "streaks" through the strong Bragg peaks are an artifact from the detector.

In two recent Letters, diffuse scattering associated with single lattice distortion (polaron), known as Huang scattering, and polaron-polaron short-range correlations have been reported in the paramagnetic phase of $(Nd_{0.125}Sm_{0.875})_{0.52}Sr_{0.48}MnO_3$ (NSSMO) [12] and $La_{1.2}Sr_{1.8}Mn_2O_7$ [13]. It is likely that the diffuse scattering at about $(0,2,0)$ in the present samples has the same microscopic origin. To confirm that the diffuse intensities are from lattice distortions and not from thermal diffuse scattering [14], we mapped out the elastic scattering around the $(0,2,0)$ reflection using the HB-1 triple-axis spectrometer with an energy resolution of $\sim 0.6$ meV FHWM. Figure 3 summarizes the outcome around $(0,2,0)$ at several temperatures for LCMO15, LCMO20, and LCMO30 manganites.

For LCMO15, the diffuse scattering shows no evidence of the butterfly shape (Fig. 3, the first row). Instead, the scattering has higher intensity along the $[0,1,0]$ direction at low temperatures (Figs. 3 and 4a). For LCMO20 and LCMO30, there are broad peaks along the wave vectors $[\pm\delta, 2\pm\delta, 0]$ in addition to the diffuse scattering. In general, the diffuse scattering arising from uncorrelated lattice distortions around the $Mn^{3+}$ ions should be doping independent because it corresponds to single polaron or Huang scattering [12,13]. The absence of the diffusive peaks along $[\pm\delta, 2\pm\delta, 0]$ in LCMO15 indicates that these peaks in LCMO20 and LCMO30 are not due to single polaron scattering but associated with polaron-polaron correlations. For LCMO20, the peaks are at $\delta \approx 0.275$ rlu from the $(0,2,0)$ and persist below $T_C$ (Fig. 3, the second row). Although similar peaks are also observed in the paramagnetic state of LCMO30 (Fig. 3, the third row), they occur at $\delta \approx 0.25$ positions and are suppressed below $T_C$. Thus, the diffuse scattering and broad peaks along $[\pm\delta, 2\pm\delta, 0]$ (Fig. 3) are probing quasi-static lattice distortions (polarons) and their short-range correlations [15]. While the result for the metallic LCMO30 is qualitatively consistent with that of NSSMO [12] and $La_{1.2}Sr_{1.8}Mn_2O_7$ [13], the modulation wave vectors along the $[\pm\delta, \pm\delta, 0]$ direction in the three-dimensional perovskites NSSMO and LCMO are rotated 45° from that of the layered compound $La_{1.2}Sr_{1.8}Mn_2O_7$, indicating different polaron (charge) ordering patterns in these two classes of materials.

To further establish the nature of the diffuse scattering, we show in Fig. 4 the evolution of the charge ordering peaks in LCMO with increasing hole doping and decreasing temperature. The scattering shows a broad peak around $(0, 2.25, 0)$ for LCMO15, while peaks for LCMO20, LCMO25, and LCMO30 are along the $[\pm\delta, 2\pm\delta, 0]$ direction. This means that the charge modulation wave vectors are along the $[0,1,0]$ direction for LCMO15 and the $[1,1,0]$ direction for LCMO20 and higher dopings. Although the microscopic origin for the ordering wave vector discontinuity is not known, we note that, at low Ca doping, the $Mn^{4+}$ ions are "impurities" causing distortions in an otherwise periodic $Mn^{3+}$ polaron lattice. Upon cooling the

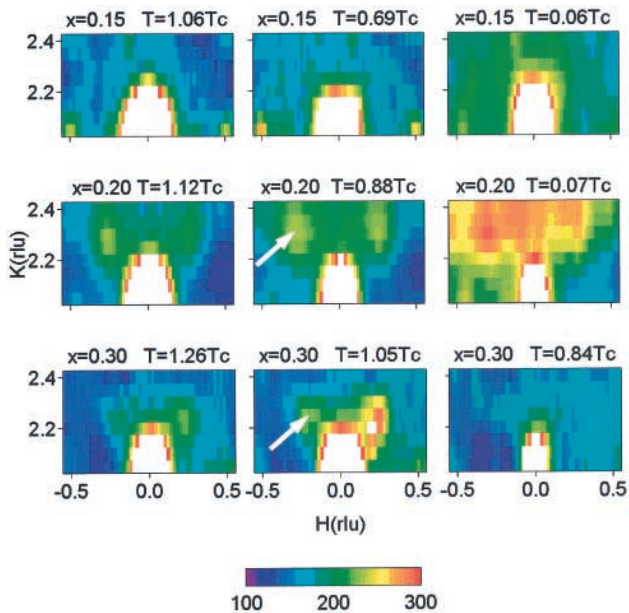

FIG. 3 (color). Diffuse scattering obtained using triple-axis spectrometers around $(0,2,0)$ fundamental reflection at various temperatures for LCMO15, LCMO20, and LCMO30. The arrows indicate the short-range charge ordering peaks in LCMO20 and LCMO30. The intensities are in arbitrary units.

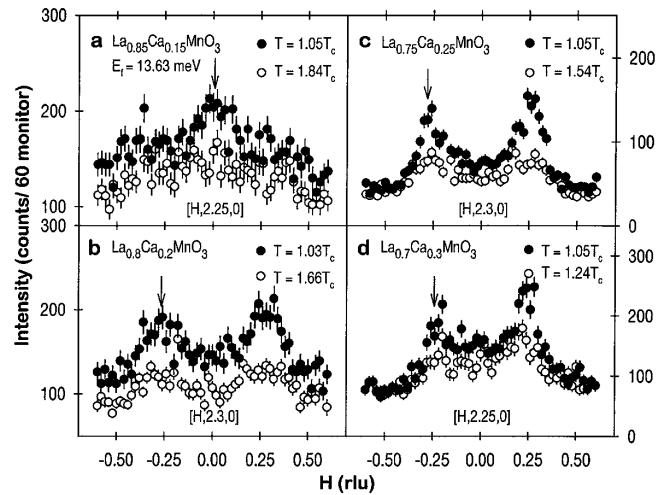

FIG. 4. Wave vector dependence of the diffuse scattering as a function of increasing doping in LCMO. The profiles in the insulating (**a**) LCMO15 and (**b**) LCMO20 are considerably broader than those of the metallic (**c**) LCMO25 and (**d**) LCMO30 at temperatures just above $T_C$, thus indicating smaller polaron coherence lengths for the insulating ferromagnets.

samples from room temperature (○) to just above their $T_C$'s (●), the profiles narrow in width and grow in intensity. For LCMO25 and LCMO30 (Figs. 4c and 4d), the coherence length of the polaron correlations increases from ~13 Å at room temperature to ~28 Å just above their $T_C$'s. The intensity of the profiles collapses below $T_C$ as a consequence of the sudden charge delocalization in the ferromagnetic metallic state [12,13]. In contrast, the short-range charge ordering peaks in the insulating ferromagnets LCMO15 and LCMO20 persist below $T_C$ and the intensities of the diffuse scattering increase with decreasing temperature (Fig. 3). Furthermore, the polaron correlation length of ~12 Å at room temperature for LCMO20 increases only slightly to ~14 Å above $T_C$ (Fig. 4b) and remains essentially unchanged in the low temperature insulating state. Thus, the insulating ferromagnets LCMO15 and LCMO20 do not develop long-range polaron or charge order as in the case of LSMO12 [8]. As a consequence, we do not expect long-range orbital order because such order is also associated with the charge ordering and $e_g$ electrons in $Mn^{3+}$ [16].

Finally, we compare the temperature dependence of the diffuse scattering with that of the resistivity on the same sample (Fig. 1). To determine how polaron-polaron correlations and single polaron scattering are related to the transport properties, we measured the temperature dependence of diffuse intensity at the charge ordering peak positions and away from them. While scattering at the ordering peak positions provides information about the population of the correlated polarons (●), the intensity away from them measures the total number of uncorrelated polarons in the system (○). For LCMO15 and LCMO30, the temperature dependent scattering at the charge ordering peak positions qualitatively follows the resistivity and is different from the single polaron diffuse scattering (Figs. 1b and 1f). Thus, the resistivity of these manganites is directly imaged by the population of the correlated polarons even though the number of uncorrelated polarons grows with decreasing temperature in the case of insulating LCMO15 but diminishes below $T_C$ for metallic LCMO30. The situation for LCMO20, on the other hand, is somewhat different. Although scattering at and away from the charge ordering peak positions exhibits very similar temperature dependence, their intensity differences become smaller at low temperatures (Fig. 1d). This suggests that the population of correlated polarons decreases with reducing temperature in LCMO20 and the resistivity rise below $T_C$ is controlled mostly by uncorrelated polarons, consistent with a polaron glass. Therefore, lattice polarons and their short-range correlations are directly imaging the transport properties of LCMO. Surprisingly, the insulating ground state in LCMO15 and LCMO20 at low temperatures does not require the development of long-range charge or orbital order. Furthermore, there are no dramatic changes in the charge ordering profiles in insulating ferromagnetic LCMO15 and LCMO20 above and below $T_C$. In contrast, LSMO12 develops long-range charge [8], orbital [9], and/or canted AF [10] order in the low temperature insulating state.

In summary, neutron scattering was used to discover the presence of lattice polarons and their short-range correlations in LCMO for the doping range $0.15 \leq x \leq 0.3$. We establish the doping dependence of the orientation, commensuration, and coherence length of the polaron correlations and show that the populations of correlated and uncorrelated polarons are intimately related to the transport properties of the materials. Our result contrasts with the expectation of the DE model, where the long-range ferromagnetic order should facilitate metallic behavior. The observation of short-range polaron correlations in insulating LCMO indicates that long-range charge [8], orbital [9], or canted AF [10] ordering are not prerequisites for localizing the conduction band electrons. Rather, the low temperature transport properties in these materials are determined by the competition between the charge correlations and the ferromagnetic DE interaction.

We thank S. Shimomura and Jiandi Zhang for helpful discussions. This work was supported by U.S. DOE under Contract No. DE-AC05-00OR22725 with UT-Battelle, LLC, by JRCAT of Japan, and by NSF DMR-9801830. The WAND was built by JAERI under the US-Japan Cooperative Program on Neutron Scattering.